\documentclass[%
superscriptaddress,
twocolumn,
showpacs,preprintnumbers,
 amsmath,amssymb,
prl,
floatfix,
]{revtex4}

\usepackage{graphicx}
\usepackage{dcolumn}
\usepackage{bm}
\usepackage{hyperref}
\usepackage{multirow}
\usepackage[usenames,dvipsnames]{xcolor}
\usepackage{color}
\usepackage{float}

\begin{document}
\newcommand\tg[1]{{\color{red}#1}}
\newcommand\tb[1]{{\color{blue}#1}}

\title{From magnetic and Kondo-compensated states to unconventional superconductivity in heavy fermions: a unified approach}

\author{Olga Howczak}
 \email{olga.howczak@uj.edu.pl}   
\affiliation{Marian Smoluchowski Institute of Physics, Jagiellonian University, Reymonta 4, 30-059 Krak\'ow, Poland}

 \author{Jan Kaczmarczyk}
 \affiliation{Marian Smoluchowski Institute of Physics, Jagiellonian University, Reymonta 4, 30-059 Krak\'ow, Poland}

\author{Jozef Spa\l ek}
\email{ufspalek@if.uj.edu.pl} 
\affiliation{Marian Smoluchowski Institute of Physics, Jagiellonian University, Reymonta 4, 30-059 Krak\'ow, Poland}
\affiliation{Faculty of Physics and Applied Computer Science, AGH University of Science and Technology, Reymonta 19, 30-059 Krak\'ow, Poland.}

\date{\today}
\begin{abstract}
Inspired by the recent experimental evidence of antiferromagnetism and superconductivity coexistence in heavy fermion CeRhIn$_5$, we propose a fully microscopic approach based on the idea of real space pairing within the Anderson-Kondo lattice model. We present an overall phase diagram incorporating the emergence of a quantum critical point, where Kondo insulating (KI), antiferromagnetic (AF) and superconducting (SC) phases meet. We also obtain the Kondo insulating state with \textit{totally compensated} magnetic moments as the parental state for the emerging SC phases. Furthermore,  the coexistent (AF+SC) phase may contain also a non-trivial spin-triplet gap component within the essentially spin-singlet pairing mechanism.      
\end{abstract}
\maketitle

The origin of \textit{unconventional superconductivity}  (\textit{superfluidity}) in strongly correlated electronic \cite{1,1a}  and optical \cite{2, 2a} lattice systems is one of the most important problems in condensed matter physics, as it concerns going beyond one of the most successful theories of XX-th century physics -\textit{ the Bardeen-Cooper-Schrieffer (BCS) theory} \cite{3} of superconductivity. One of the novel pairing mechanisms is based on the Dirac's \textit{ universal idea of exchange interaction} \cite{6}. This idea has a direct application in the case of \textit{ high temperature superconductivity}, where it takes the form of the  \textit{$t-J$ model with real space pairing} \cite{7,7a, 8}. Here we extend these  ideas  to \textit{the heavy-fermion systems} \cite{11} and show that it leads to a unique connection between antiferromagnetism (AF) and superconductivity (SC), in accordance with the recent experimental results \cite{17,17a}. Also, the evolution from the Kondo insulating state (PKI) via pure d-wave SC to the coexistent AF+SC phase, leads to highly non-trivial structure of the order parameters.

 In this Letter we present a fairly complete  phase diagram and thus demonstrate in a straightforward manner the applicability of the concept of real-space pairing  to this very important class of strongly correlated quantum materials. We also show that the system properties can evolve from either PKI  or antiferromagnetic state to the unconventional SC paired states. In this manner, we  extend the universal meaning of the concept of real-space pairing in the strongly correlated systems \cite{11, 12, PhysRevB.84.064514, 12a}. Also, in the PKI state the magnetic moments of $f$ electrons and conduction ($c$) electrons  \textit{totally compensate each other}. This last result provides a viable and intuitively appealing definition of the Kondo insulators and singles out their difference with either Mott insulators or intrinsic semiconductors. 
Furthermore, the paired coexistent antiferromagnetic-superconducting (AF+SC) state  contains also a spin-triplet component within the essentially spin-singlet pairing mechanism \cite{14, 14a}. All of these results,  have been obtained for the first time and within a single formal approach. 

The results are discussed starting from  introduced by us  \textit{Anderson-Kondo lattice Hamiltonian} \cite{11,11a}, in which both the Kondo interaction and the residual hybridization processes appear together and are accounted for on an equal footing. Subsequently, a modified Gutzwiller approach is formulated \cite{13a} and combined with the  \textit{statistically consistent renormalized mean-field theory} (SC-RMFT) \cite{15,10}. In effect we show that \textit{the hybrid} ($f$-$c$) type of real space  pairing induced by the Kondo-type interaction is crucial for description of heavy-fermion SC properties even though the $f$-$f$ interaction may play the dominant role in magnetism of $f$-electron system in the localization limit. We also discuss at the end the role of $f$-$f$ intersite interaction to demonstrate the stability of our solution with respect to this higher-order interaction.     

In the case of cerium compounds, one usually assumes that: (i) only the $\Gamma_7$ doublet ($4f^1$) state of Ce$^{+4-n_f}$ valency, with the $f$-level occupancy  $n_f\rightarrow 1$   in the low-temperature dynamics; (ii) this state is strongly hybridized with the conduction-band ($c$) states; and (iii) the hybridization has either intraatomic or interatomic form. We discuss here only the interatomic case, as it leads to the stable pairing gap of d-wave character observed almost universally in strongly correlated systems \cite{1a}. Formally,  in the limit of large intraatomic $f$-$f$ Coulomb interaction of magnitude $U$ the  effective  Hamiltonian in the real-space representation  
has the following form:
\begin{widetext}
\begin{eqnarray}
\mathcal{H} &=&  
\sum_{mn\sigma} \left(t_{mn} c^{\dagger}_{m\sigma} c_{n\sigma} -\sum_{i} \frac{V_{im}^{*} V_{in}}{U + \epsilon_f} \hat{\nu}_{i\bar{\sigma}}c^{\dagger}_{m\sigma} c_{n\sigma}  \right) + 
\sum_{i\sigma}{\epsilon_f } \hat{\nu}_{i\sigma} + \sum_{imn \sigma} \frac{ V_{im}^{*} V_{in} }{U + \epsilon_f}\hat{S}_{i}^{\sigma} c^{\dagger}_{m\bar{\sigma}} c_{n\sigma} \nonumber\\
 &+&  \sum_{im\sigma} (1-\hat{n}^f_{i\bar{\sigma}}) (V_{im} f^{\dagger}_{i\sigma} c_{m\sigma} +  H.c.)
+
\sum_{im} J^K_{im} \left(\mathbf{\hat{S}}_i\cdot \mathbf{\hat{s}}_m - \frac{\hat{\nu}_{i}\hat{n}^c_{m}}{4}  \right) + \sum_{ij} J^{H}_{ij} \left(\mathbf{\hat{S}}_i\cdot \mathbf{\hat{S}}_j - \frac{\hat{\nu}_{i} \hat{\nu}_{j}}{4} \right) \nonumber\\
&-&
 \frac{1}{2} g\mu_B H \sum_{i\sigma} \sigma \hat{\nu}_{i\sigma} - \frac{1}{2} g\mu_B H \sum_{m\sigma} \sigma \hat{n}^c_{m\sigma},
\label{eq:1}
\end{eqnarray} 
\end{widetext}
with $J^K_{im}\equiv 2|V_{im}|^2/(U+\epsilon_f)$ and $J^H_{ij}\equiv\sum_{mn}|V_{im}V_{jn}|^2/(U+\epsilon_f)^3$.
The consecutive terms represent the following dynamical processes: the first comprises a direct ($c$-$c$) hoping in the conduction band, as well as the hoping via intermediate $f$-state. The second and the third express, respectively, the bare $f$-electron energy  and the spin-flip term. The fourth term contains the residual hybridization. The fifth and the sixth term represent, respectively, the Kondo ($f$-$c$) interaction  and the Heisenberg ($f$-$f$) term (in both cases full Dirac-exchange operators are taken).  The last one represents the Zeeman term for both $f$ and $c$ electrons. Here we have projected out completely the double occupancies of the $f$ states, what is equivalent to assuming that \textit{$f$ electrons are strongly correlated}. In effect the $f$-electron number $\hat{n}^f_{i\sigma}$ is replaced  by their projected counterpart: $\hat{\nu}_{i\sigma} \equiv \hat{n}^f_{i\sigma}(1-\hat{n}^f_{i\bar{\sigma}})\equiv  \tilde{f}^\dagger_{i\sigma}\tilde{f}_{i\sigma}$ and $\hat{\nu}_{i}\equiv \sum_{\sigma}\hat{\nu}_{i\sigma}$. The $f$-spin operator is defined by $\mathbf{\hat{S}}_i\equiv (\hat{S}_i^{\sigma}, \hat{S}^{z}_i )\equiv [\tilde{f}^{\dagger}_{i\sigma} \tilde{f}_{i\bar{\sigma}}, 1/2 (\hat{\nu}_{i\uparrow}-\hat{\nu}_{i\downarrow})]$. The corresponding (unprojected) quantities for $c$-states are $\hat{n}^c_{m\sigma}$, $\hat{n}^c_{m}$ and  $\mathbf{\hat{s}}_m$.
To discuss the system properties  a number of quantities  are first to be calculated self-consistently: the $f$-level and $c$-band occupancies: $n_{f,c}\equiv\sum_{\sigma} \langle \hat{n}^{f,c}_{i\sigma}\rangle$; the $f$- and $c$-magnetic moments: $m_{f,c}\equiv \sum_{\sigma}\sigma \langle \hat{n}^{f,c}_{i\sigma}\rangle$; the hybridization correlation and  the hybrid pairing amplitudes: $\gamma_\sigma\equiv \langle f^\dagger_{i\sigma} c_{m\sigma}\rangle$ and   $\Delta_\sigma\equiv \langle f_{i\sigma} c_{m\bar{\sigma}}\rangle$, the $f$-$f$ and $c$-$c$ hopping correlations $\chi_{\sigma}\equiv \langle f^\dagger_{i\sigma} f_{j\sigma} \rangle$ and $\xi_{\sigma}\equiv \langle c^\dagger_{n\sigma} c_{m\sigma} \rangle$, as well as the chemical potential $\mu$. 

To calculate the averages of operators (e.g. $\langle f^\dagger_{i\sigma} c_{m\sigma}\rangle $) we utilize the extended Gutzwiller scheme \cite{13a}.  We also include the magnitude of the intersite spin-singlet $f$-$f$ pairing $\Delta_{ff}\equiv \langle f_{i\uparrow} f_{i\downarrow}\rangle$ and discuss it separately below. In addition to those, we have to introduce and calculate the molecular fields appearing through \textit{statistical consistency conditions} so that  the self-consistent equations for the above quantities coincide with those calculated variationally \cite{11a}. 
Our microscopic model is characterized then by the total  number of electrons $n_e = \langle\hat{n}^f_{i\sigma}\rangle+\langle\hat{n}^c_{i\sigma}\rangle$; the
magnitude $V<0$ of the $\mathbf{k}$-dependent hybridization $V_{\mathbf{k}}=8(|V|/W)\xi_{\mathbf{k}}$, bandwidth $W$, and  the bare conduction-band dispersion relation $\xi_\mathbf{k}=-(W/4)(\cos(k_x)+\cos(k_y))$, (i.e., for exemplary two-dimensional lattice, applicable to systems such as CeMIn$_5$, with M=Co, Rh, Ir); the position $\epsilon_f$ of the bare $f$ level; and the magnitude $U$ of the $f$-$f$ interaction. The parameters $\epsilon_f$, $V$, $U$, and $n_e$ are independent; also, all the energies are taken in units of $W$, wherever not specified explicitly. 
 \begin{figure}
  \centering
	\includegraphics[height=0.45\textwidth, angle=-90]{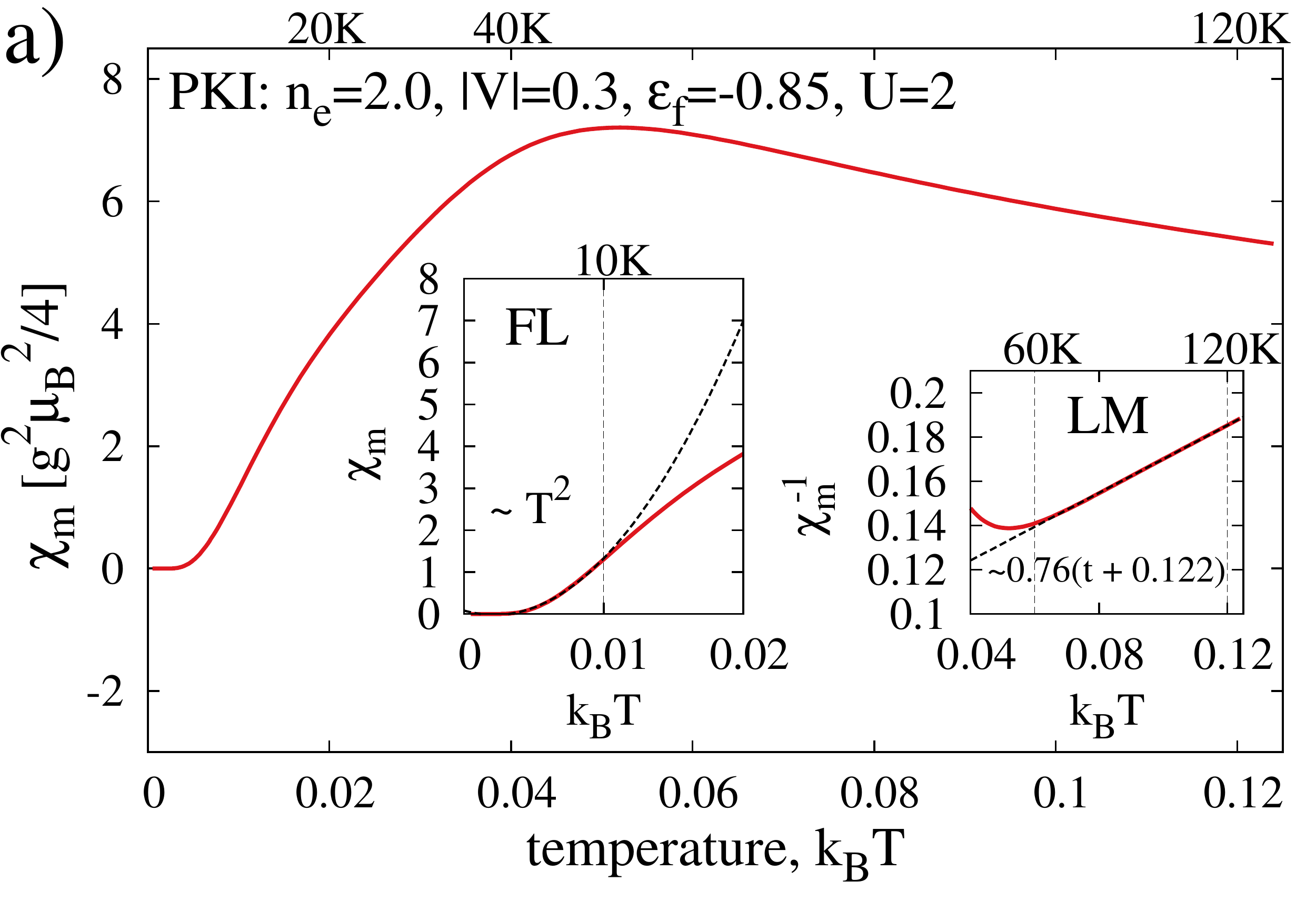}
	
	\includegraphics[height=0.45\textwidth, angle=-90]{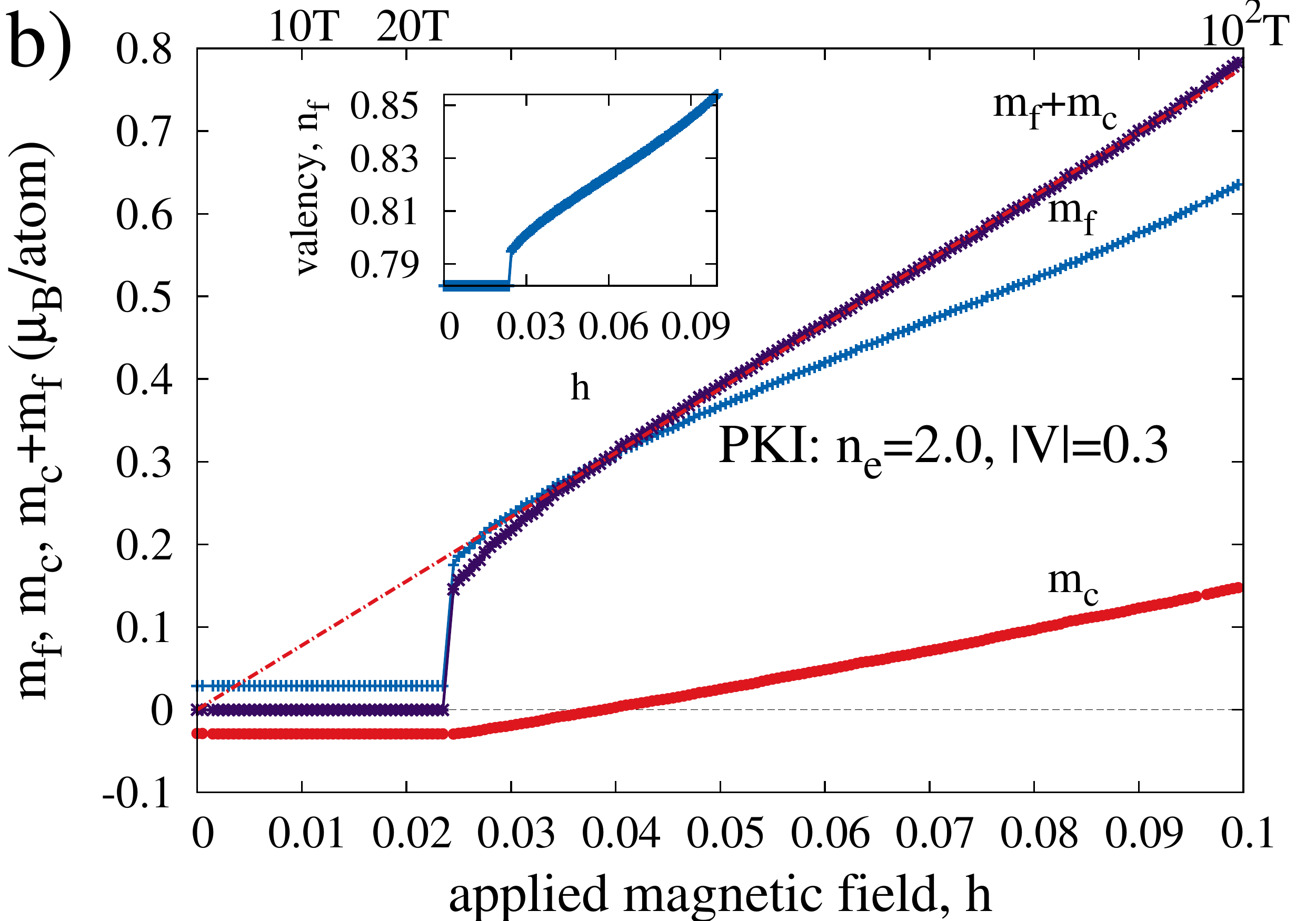}
	\caption{(color online). a): temperature dependence of the zero-field magnetic susceptibility for the nonmagnetic (parent) Kondo insulator (PKI). {The insets show PKI evolution from the Fermi-liquid regime (FL), to the localized moment regime (LM)}. The upper temperature scale is provided for bare $c$ bandwidth $W=10^3K$; b): applied field $h\equiv g\mu_BH/(2W)$ dependence of magnetic moments and of the $f$-level occupancy (inset).}
\label{fig:1}
 \end{figure}

We discuss first the reference Kondo-compensated (PKI) state. In Fig. \ref{fig:1}a we display the temperature dependence of the spin susceptibility $\chi_{m}$. The crucial point is that the spin susceptibility  $\chi_{m}(T=0)\equiv 0$, as one would expect for the totally compensated-moment KI state with a small gap or pseudogap. However, in distinction to the semiconductors, the system exhibits $T^2$ behavior at low temperature charactering the Fermi-liquid state. Also, the $\chi_{m}(T)$ exhibits a maximum for $T_{max} = 50 \div 250K$  (when $W$ is in the range $\sim0.1-0.5eV$)
 followed by the Curie-Weiss behavior with a large (and of AF sign) paramagnetic  Curie  temperature in the range  $\Theta=100-150K$. These features, together with the universal scaling law $\chi_m\rho=const.$, with $\rho$ being the system resistivity, proposed and tested earlier \cite{16}, provide us with the confidence that our method of approach describes properly the universal trend of the experimental data for KI systems \cite{16}.
Furthermore, we have shown Fig. \ref{fig:1}b the applied magnetic-field dependence of the $f$, $c$, and the total magnetic moments at $T=0$. The total moment remains completely compensated in low fields and \textit{this is a true Kondo-lattice state}, broken at the metamagnetic point $\mu_B H_m/W\simeq0.02$. The magnetization curve above metamagnetic critical field $H_m$ shows approximately linear behavior, $m_{t}\sim H$, as marked. In the inset we show field dependence of the $f$-level occupancy (valency). Note that even though the ground state is insulating, the  valency deviation from Ce$^{3+}$ configuration is $1-n_f=\delta\sim 0.2$ i.e., is not integer ($n_f=1$ case  corresponds to the $Ce^{3+}$ localized-moment state and is reached only as $|V|\rightarrow0$).
  \begin{figure}
  \centering
	\includegraphics[width=0.5\textwidth]{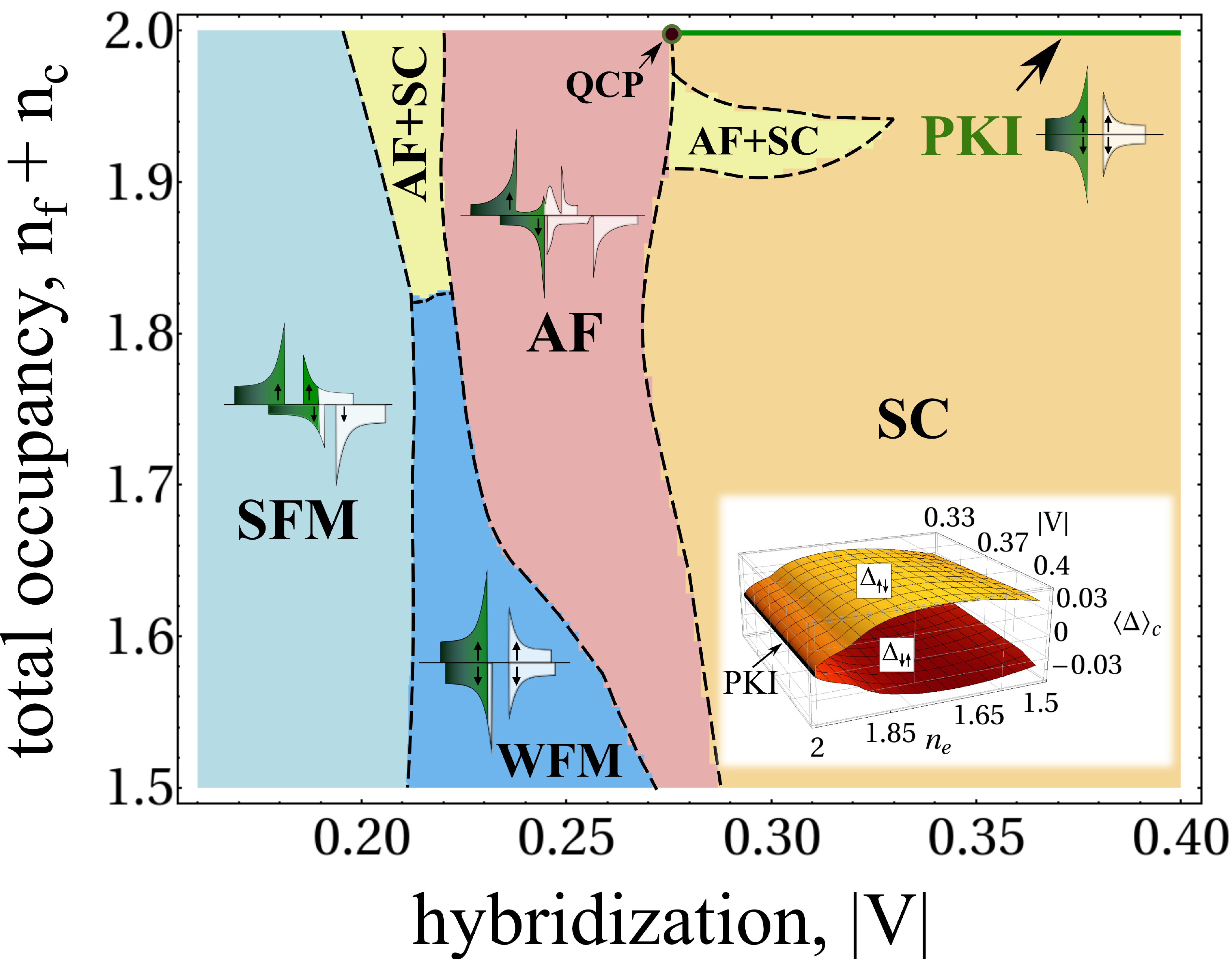}
	\caption{(color online). Overall phase diagram on the plane number of electrons $n_e\equiv \langle n^f_i\rangle+\langle n^c_i\rangle$ - interatomic hybridization magnitude  $|V|$, comprising strong ferromagnetic (SFM), weak ferromagnetic (WFM), coexistent (AF+SC),  antiferromagnetic (AF),  superconducting (SC), and the parent Kondo insulating  (PKI) states. Inset: evolution of the SC state from PKI; the spin components of $\langle\Delta\rangle_c$ the spin-singlet gap parameter: $\langle\Delta_{\uparrow}\rangle_c\equiv \Delta_{\uparrow\downarrow}\equiv\langle f_{i\uparrow}c_{m\downarrow} \rangle_c$ and $\langle\Delta_{\downarrow}\rangle_c\equiv \Delta_{\downarrow\uparrow}\equiv\langle f_{i\downarrow}c_{m\uparrow} \rangle_c$  are shown.  The solid dot marks the AF metal - Kondo insulator (PKI) quantum critical point (QCP). The parameters are $U=2$, $\epsilon_f=-0.85$. The representative density-of-states curves  define the normal-phase distinctions.}
\label{fig:3}
 \end{figure}
 
\begin{figure}
  \centering  
 \includegraphics[height=0.5\textwidth, angle=-90]{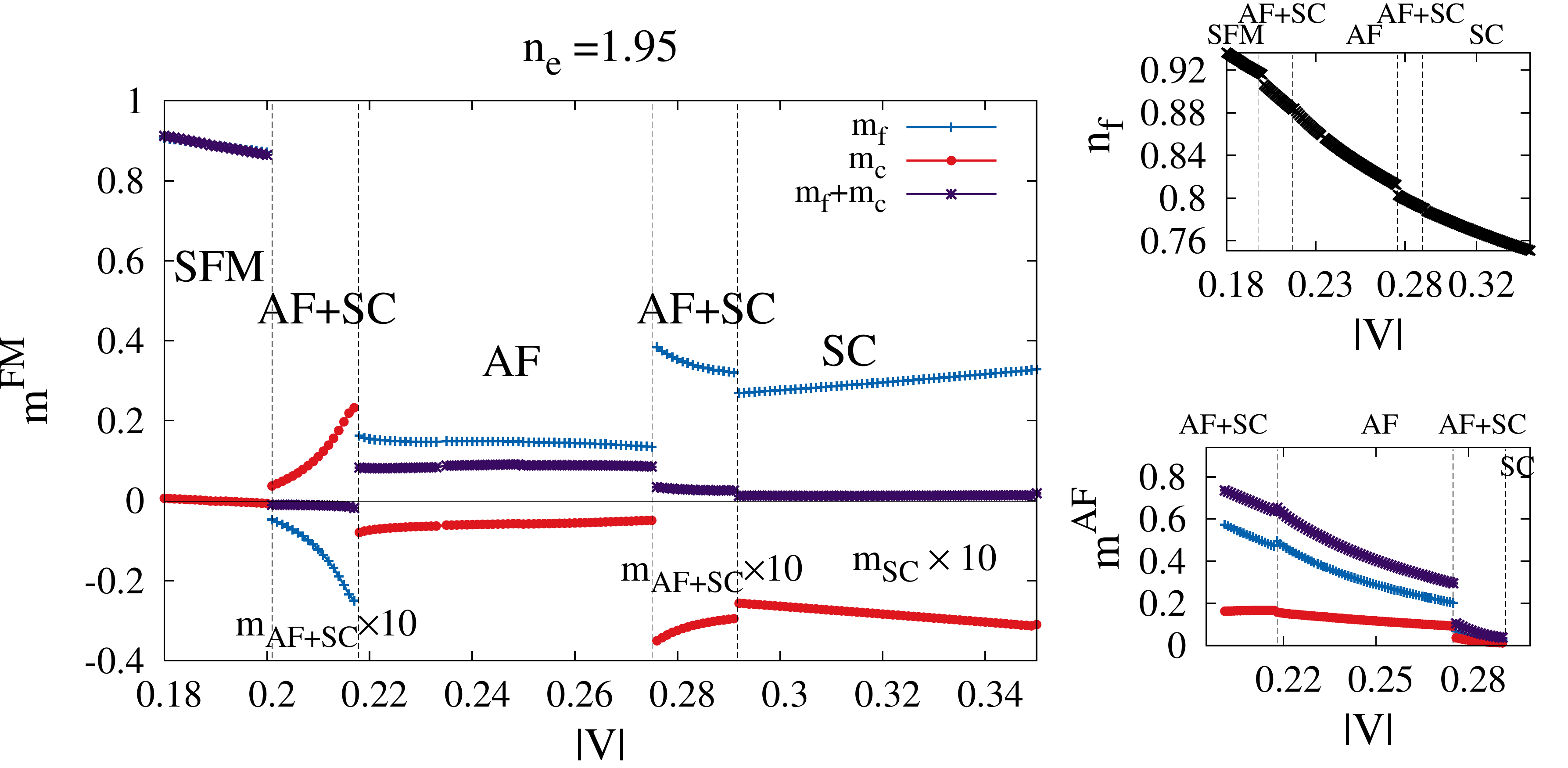} 
  \caption{(color online). Magnetic moments as a function of $|V|$. The left panel presents the uniform  component  $m^{FM}$ of magnetization in the specified  phases.  The top right  panel shows the $f$-level occupancy $n_f$ as a function of the hybridization and the bottom  represents the staggered component of magnetization, $m^{AF}$. The magnitude of the uniform components of magnetic moments was multiplied by the factor 10 in the AF+SC and 
   SC phases.}
  \label{fig:4}
\end{figure}
 \begin{figure}
  \centering
	\includegraphics[width=0.3\textwidth]{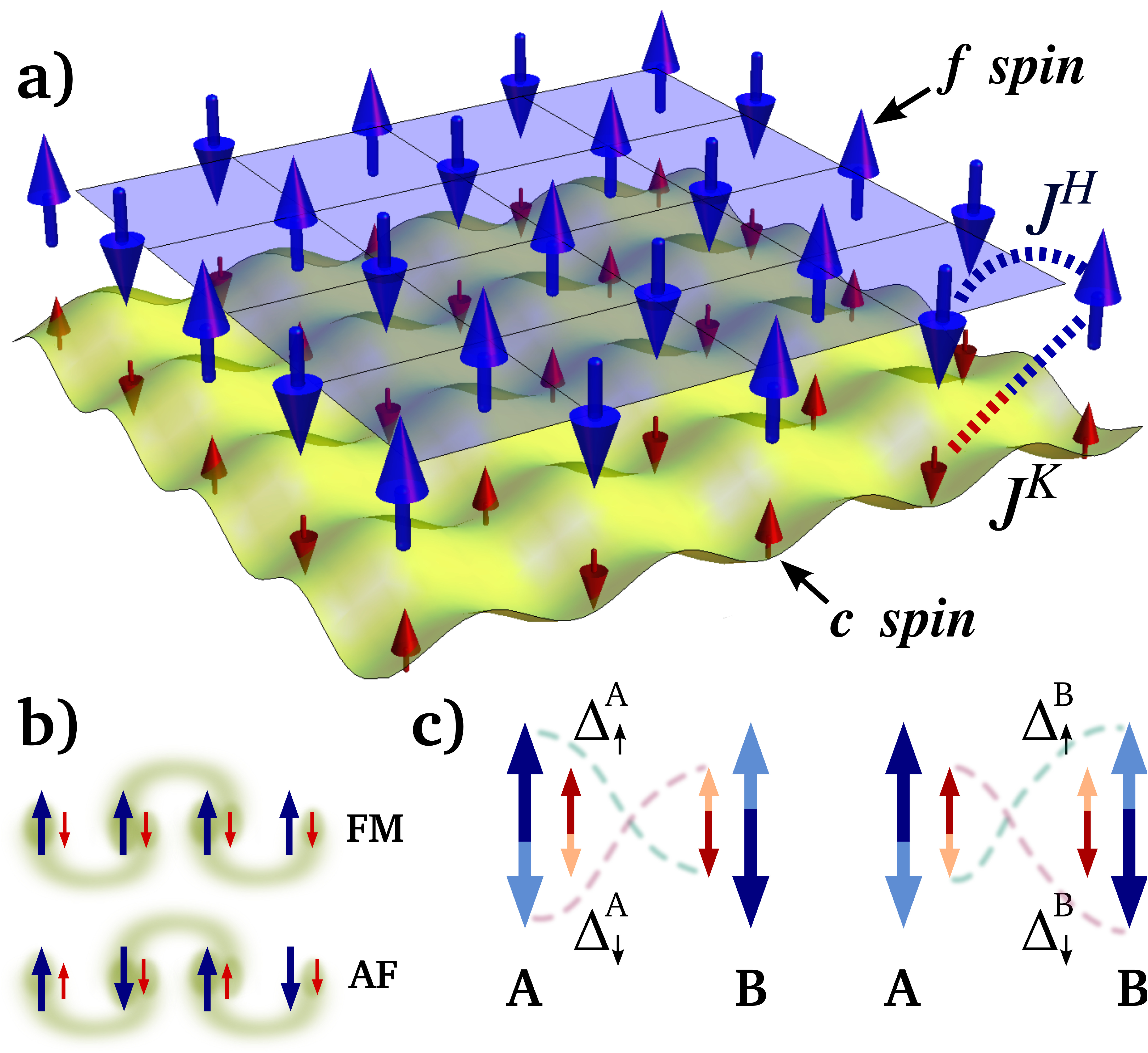}
	\caption{(color online). a) illustration of the two different AF orders as a consequence of both the $f$-$f$ ($\propto J^H$) and Kondo-type $f$-$c$ ($\propto J^K$) spin-spin interactions; b) orientation of the $f$-moments (blue) and the $c$-moments (red) in FM and AF phases, respectively; c) different components of the hybrid ($f$-$c$) superconducting gap in the AF+SC phases, which compose the spin-singlet $\Delta^S_{\sigma}\equiv(\Delta^A_{\sigma}+\Delta^B_{\sigma})/2$  and the spin-triplet $\Delta^T_{\sigma}\equiv(\Delta^A_{\sigma}-\Delta^B_{\sigma})/2$ components of the  gap. }
	  \label{fig:5}
 \end{figure} 

 \begin{figure}
  \centering

	\includegraphics[height=0.5\textwidth, angle=-90]{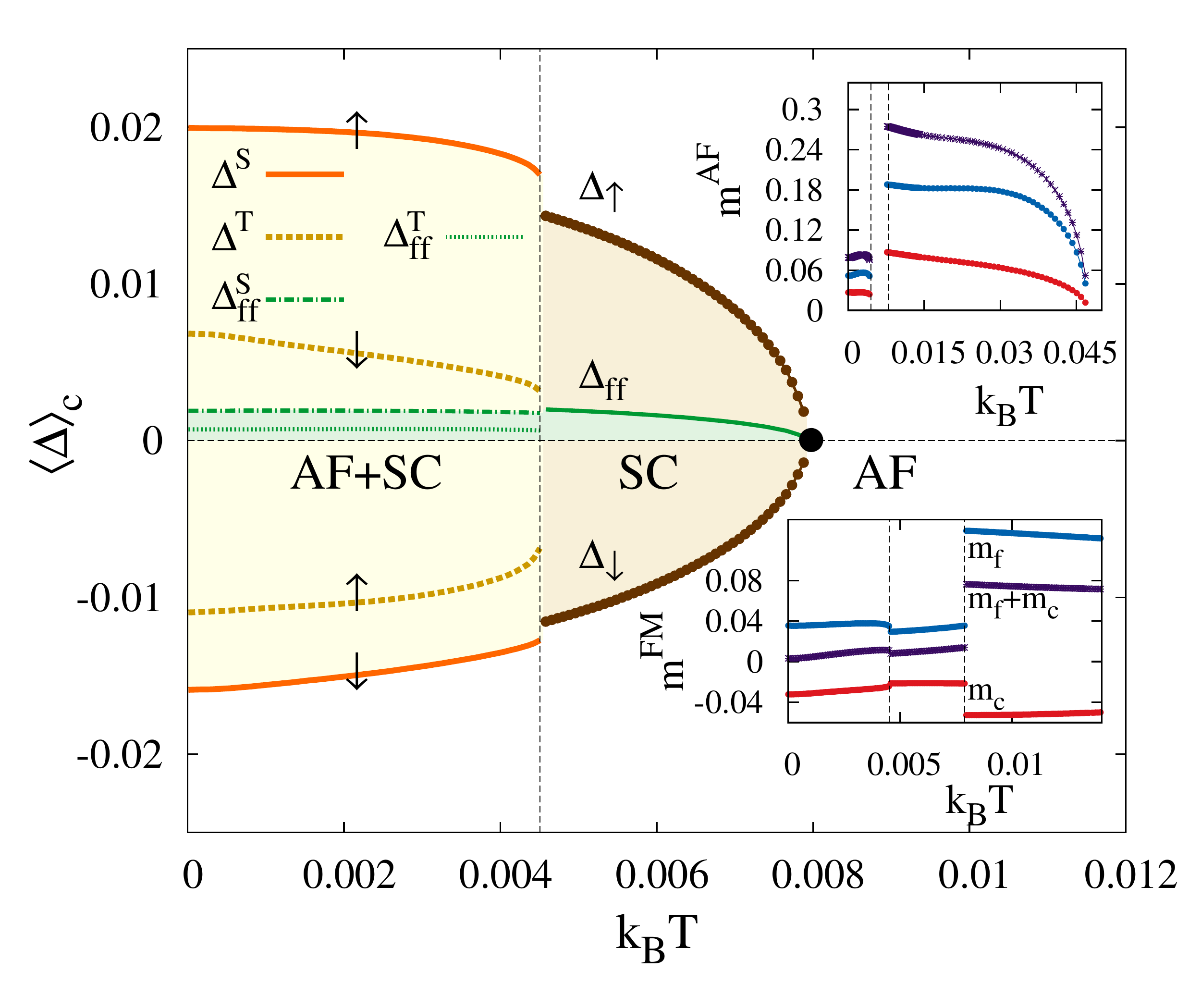}
	
	\caption{(color online). Temperature dependence of the correlated-gap components, as well as of  $m^{AF}$ and $m^{FM}$ (insets). The parameters: $n_e=1.95$, $|V|=0.28$.  The arrows in the AF+SC phases label $\Delta_\sigma^{S,T}$ components with $\sigma = \uparrow$ or $\downarrow$.}
\label{fig:6}
\end{figure}  
 
We turn now to the analysis of magnetic and superconducting phases appearing in the metallic state, i.e., with the partial filling of orbitals, $n_e<2$. In Fig. \ref{fig:3} we draw an overall phase diagram on the plane occupancy $n_e$ - hybridization magnitude $|V|$. This diagram contains two coexisting  AF+SC phases in the heavy-fermi-liquid regime ($n_e\rightarrow 2$), as well as  AF and SC phases. In the small-hybridization limit we encounter two ferromagnetic phases (WFM and SFM). In insets of the magnetic part we characterize the  phases by the corresponding shapes of the quasiparticle densities of states and the position of the chemical potential. Also, in the lowest right inset we show the evolution of the \textit{correlated hybrid gap} $\langle\Delta\rangle_c$ components $\langle\Delta_\uparrow\rangle_c \equiv \Delta_{\uparrow\downarrow}\equiv \langle f_{i\uparrow}c_{m\downarrow}\rangle_c$ and  $\langle\Delta_\downarrow\rangle_c \equiv \Delta_{\downarrow\uparrow}\equiv \langle f_{i\downarrow}c_{m\uparrow}\rangle_c$. They are of equal amplitude and of opposite sign only when the superconducting state is of pure spin-singlet nature. The phase diagram drawn here along the line $n_e\sim 1.95$ reflects the sequence of phases observed recently in CeRhIn$_5$ (AF$\rightarrow$AF+SC$\rightarrow$SC) with the increasing pressure \cite{17}.
In general, as in the case of high temperature superconductors, pure SC state evolves with the diminishing electron count from the insulating, here nonmagnetic PKI state. Also, in distinction to the high-T$_c$ systems,  we have here two AF kinetic-exchange interactions:  the Kondo $f$-$c$ and  $f$-$f$ type, both arising from the virtual inter-band $c\rightleftarrows f$ hopping. 
To characterize the phases marked in Fig. \ref{fig:3}, we have drawn in Fig. \ref{fig:4}  evolution of the calculated \textit{uniform} ($m^{FM}$) and the \textit{staggered} ($m^{AF}$) components of the  magnetic moment, defined through the relation $n_{f(c)i\sigma} \equiv \langle \hat{n}^{f(c)}_{i\sigma}\rangle_{0} \equiv \frac{1}{2} (n_{f(c)}+ \sigma m^{FM}_{f(c)} + \sigma m^{AF}_{f(c)} e^{i\mathbf{QR_i}})$, with $\mathbf{Q}=(\pi,\pi)$ being the AF-ordering wave vector and $\mathbf{R_i}$ - the lattice-site position. Fig. \ref{fig:4} conveys our next principal messages coming out from our fully microscopic theory. First, the uniform moment component  is practically  compensated in the AF+SC phase (the exemplary numerical values are listed in table 1 of \textit{Supplementary Material}). Second, there appears a very small  staggered moment in the second to-the-right AF+SC state. This may explain, why it may be difficult to measure it in the coexistent phase as discussed by Knebel \textit{et al.} \cite{17}. Third, most of the transition lines mark discontinuities and thus, their first-order nature as a function of $|V|$ (emulating the changing  pressure applied to the system). Fourth, the coexistent AF+SC phases border with AF and SC states and have a very small uniform magnetic moment. In general, the diagonal (spin-spin) correlations lead to the Kondo compensation and the off-diagonal (pairing) correlation lead to the emergence of superconducting state. Finally, the valence of Ce$^{+4-n_f}$ evolves systematically from that for heavy-fermion $n_f=1-\delta$, with $\delta\ll 1$, to that characterizing the fluctuation-valence regime ($n_f<0.75$), with the increasing hybridization.

Before characterizing the temperature evolution of the  gap components, we sketch in Fig. \ref{fig:5} both configurations of the component moments in AF  and FM phases, as well as define different  gap components (c), that we have to introduce in order  to describe the coexistent AF+SC phases. Such multiplicity of the gaps is enriched when we include also the $d$-wave gap coming from the $f$-$f$ pairing.  It is important to note that the spin-triplet component appears only in the coexistent AF+SC phase \cite{14a}, as there are majority and minority spins then, e.g. on sublattice A  we have that $\langle\hat{n}^f_{i\uparrow}\rangle > \langle\hat{n}^f_{i\downarrow}\rangle$ and similarly for the $c$ electrons;  hence $\Delta^A_{\uparrow}\neq \Delta^B_{\uparrow}$.
In  Fig. \ref{fig:6} we show that all the gaps define still a single transition temperature for given phase. One should note the sequence AF+SC$\rightarrow$SC$\rightarrow$AF$\rightarrow$paramagnet. AF$\rightarrow$paramagnet transition temperature is of an order of magnitude higher than that for AF+SC$\rightarrow$SC transition. The $f$-$f$ pairing coming from the fourth-order contribution of magnitude $J^H_{\langle ij\rangle}=12V^4/(\epsilon_f+U)^3$ is also taken into account in this panel and, in distinction to high-T$_c$ systems, here the hybrid ($f$-$c$) pairing contribution is dominant. In all the phases we have that $n_f\sim 0.9\div0.8$.

To recapitulate, the strong correlations, crucial for the appearance of both the AF and the Kondo-compensated states, are responsible to an equal extent for the emergence of both the coexistent AF+SC and the pure SC phases. The sequence of the phases AF$\rightarrow$AF+SC$\rightarrow$SC appears with the increasing $V/\epsilon_f$ ratio, as observed in quasi-two-dimensional CeRhIn$_5$ \cite{17, 17a,19}. Our real-space pairing thus accommodates in a natural manner into the spin  correlations in AF phase (cf. Fig. \ref{fig:5}c). The interesting feature is also the presence of a very weak  staggered component of magnetic moment in the second AF+SC phase (cf. Fig. \ref{fig:4}). The principal result is also the strong mutual $f$- and $c$-moment compensation, either complete or partial, depending on the phase. Furthermore, the phenomenon assisting the uniform moment is the appearance of the spin-triplet gap component  in the coexistent AF+SC phases. 

The authors acknowledge the support of the Polish Foundation for Science (FNP) through Grant TEAM (2011-14). O.H. has been also supported by the Ministry of Science and Higher Education, Grant No. NN202 489839. The authors are grateful to Prof. Piers Coleman for a discussion.


\end{document}